\documentclass{article}
\usepackage{amsmath}
\usepackage{amssymb}
\usepackage{graphicx}
\begin{document}
\begin{center}
\Large
\textbf{Quantum transition probability} 

\textbf{in convex sets and self-dual cones}
\vspace{0,3 cm}
\normalsize

Gerd Niestegge 
\footnotesize
\vspace{0,2 cm}

Ahaus, Germany

gerd.niestegge@web.de, https://orcid.org/0000-0002-3405-9356
\end{center}
\normalsize
\begin{abstract}
The interplay between the algebraic structure 
(operator algebras) for the quantum
observables and the convex structure of the state space 
has been explored for a long time and most advanced results 
are due to Alfsen and Shultz. Here we present a 
more elementary approach with a more general structure
for the observables,
which focuses on the transition 
probability of the quantum logical atoms.
The binary case gives rise to the generalized qubit models
and was fully developed in a preceding paper.
Here we consider any case with finite information capacity
(binary means that the information capacity is 2).
A novel geometric property that makes any
compact convex set a matching state space
is presented. Generally, the 
transition probability is not symmetric; if it is symmetric,
we get an inner product and a self-dual cone. 
The emerging mathematical structure comes close to 
the Euclidean Jordan algebras and becomes 
a new mathematical model for a potential extension of quantum theory.
\vspace{0,3 cm}

\end{abstract}

\section{Introduction}

The study of the interplay between the algebraic structure
for the quantum observables and the convex structure 
for the quantum state space has a long history.
Most advanced results are due to Alfsen and Shultz,
who began their work in the 1970ies and later 
wrote an updated and complete presentation 
of their results in two monographs \cite{AS01, AS02}.
They succeeded in characterizing those convex sets 
that are the state spaces of the
$C^{\ast}$-algebras, von Neumann algebras 
and the Jordan analogues of these operator algebras
(the JB and JBW algebras).

Here we present a different more elementary approach,
starting from minimal assumptions for the system of observables
and then focusing on the transition probability
of the quantum logical atoms
as defined in Refs. \cite{nie2020alg_origin, nie2021generic, Nie2022genqubit}.
The compressions which are an important ingredient
in Alfsen and Shultz's theory 
are not used and do not generally exist here.
Our approach is more general - at least
in the finite dimensional case.
In the infinite dimensional case, 
Alfsen and Shultz's theory 
includes the non-atomic type II and type III
von Neumann algebras \cite{AS01},
which are not covered by our approach.

The binary type of our approach
(the generalized qubit models) 
was fully developed in a preceding paper \cite{Nie2022genqubit},
where the associated state spaces turned out to be 
the strictly convex and smooth compact convex sets. 
Here we extend this model to cover more general types 
and not only the binary one.
A novel geometric property
that makes any compact convex set a matching state space
is presented. 
This property has a purely mathematical character, 
but it is shown that, in many cases,
it becomes equivalent to
two better interpretable and physically more plausible
postulates: spectrality and strongness of the state space.

Generally the transition probability need not be symmetric;
its symmetry results in an inner product in the order unit space, 
making it a real Hilbert space with a self-dual cone. 
The reverse task is also tackled and two properties
are identified that impose on any 
finite dimensional real Hilbert space with a self-dual cone
the desired structure with symmetric transition probability.

We do not reconstruct quantum theory, 
but come close to it. Our model
can be considered a mathematical structure for an
extension of quantum theory, which would involve
some interesting implications on physics - 
particularly on the theoretical foundations of quantum measurement.

We here apply the mathematical formalism of order unit spaces and their state spaces \cite{AS01,AS02}
and enter it directly. The same formalism plays a central role in the generalized probabilistic
theories \cite{barnum2013ensemble, barrett2007information, muller2021probabilistic, plavala2023general}, 
where it is usually derived from a certain collection of operational postulates.

The paper is organized as follows.
Some axioms and properties of an order unit space,
which are required to define and ensure the existence
of the transition probabilities, 
are introduced in section 2.
In section 3, we turn to the compact convex sets 
and the geometric property that makes them 
state spaces.
Finiteness of the information capacity is introduced 
and studied in section 4. This condition is 
needed in section 5, where the relationship
between spectrality, strongness of the state space
and the geometric condition is elaborated.
The consequences of the symmetry of the transition probability
are analyzed in section 6; these are an inner product and a self-dual cone.
In section 7, we address the reverse question of what properties 
would impose the structure studied in sections~2~-~6 on any finite
dimensional real Hilbert space (Euclidean space) with a self-dual cone.
Some physically important implications of our approach,
if it were used as a mathematical model
for an extended quantum theory, are discussed in section~8.

\section{Minimal extreme points}

As in the preceding paper \cite{Nie2022genqubit}, 
our mathematical structure for the observables
shall be an order unit space $A$. Note that
the self-adjoint operators of customary quantum mechanics,
the self-adjoint parts of the von Neumann algebras and the JBW algebras
form such spaces and that the associated quantum logics are
the extreme boundaries of the positive parts of the unit balls \cite{AS02, hanche1984jordan}.

An order unit space $A$ possesses an order relation $\leq$, a distinguished
order unit $\mathbb{I}$ and order norm $\left\| \  \right\|$ \cite{AS01, AS02, hanche1984jordan}.
The norm is connected to the order unit via 
$\left\| a \right\| = inf \left\{s > 0 \ | \  -s \mathbb{I} \leq a \leq s \mathbb{I} \right\}$
for $a \in A$.
The extreme boundary of the unit interval 
$\left[0,\mathbb{I}\right] := \left\{a \in A: 0 \leq a \leq \mathbb{I} \right\}$ $\subseteq A$
is denoted by $ext(\left[0,\mathbb{I}\right])$.
Obviously $0,\mathbb{I} \in ext(\left[0,\mathbb{I}\right])$.
\vspace{0,3 cm}

\noindent
\textbf{Lemma 1:} 
\itshape
If $p \in ext(\left[0,\mathbb{I}\right])$, 
then $\mathbb{I} - p \in ext(\left[0,\mathbb{I}\right])$.
\vspace{0,3 cm}
\normalfont

Proof. Suppose $\mathbb{I} - p = sa + (1-s)b$ 
with $p \in ext(\left[0,\mathbb{I}\right])$, 
$a,b \in \left[0,\mathbb{I}\right] $ 
and $s \in \mathbb{R}$, $0 < s < 1$. Then 
$p = s(\mathbb{I} - a) + (1-s)(\mathbb{I} - b)$.
Since $p$ is an extreme point of $\left[0,\mathbb{I}\right] $, we get 
$\mathbb{I} - a = \mathbb{I} - b = p$ and thus $a = b = \mathbb{I} - p$.
Therefore $p$ is an extreme point 
in $\left[0,\mathbb{I}\right] $. \hfill $\square$
\vspace{0,3 cm}

A \emph{state} is a positive linear functional $\mu : A \rightarrow \mathbb{R}$
with $\mu(\mathbb{I}) = 1$. The states form a $w^{\ast}$-compact subset $S_A$
of the dual of $A$; $S_A$ is called the \emph{state space} of $A$
and its extreme points are the \emph{pure states}.

An element $e \in ext(\left[0,\mathbb{I}\right])$ with $e \neq 0$ 
is called a \emph{minimal} extreme point
if there is no $p \in ext(\left[0,\mathbb{I}\right])$ 
with $p \leq e$ and $0 \neq p \neq e$.
Having in mind the results from the preceding paper \cite{Nie2022genqubit}
(or the situation in the von Neumann algebras and the JBW algebras),
the system $ext(\left[0,\mathbb{I}\right])$
is the candidate for the \emph{quantum logic} and
the minimal extreme points will then become the \emph{atoms}. 

The norm of each non-zero element $p$ 
in $ext(\left[0,\mathbb{I}\right])$ is $\left\|p\right\| = 1$;
therefore there is a state $\mu$ with $\mu(p)=1$. 
The uniqueness of this state in the case when 
$p$ is a minimal extreme point becomes our first postulate.
It is motivated by the requirement
that the \emph{transition probability} defined 
in Refs. \cite{nie2020alg_origin, nie2021generic, Nie2022genqubit}
shall exist for the atoms.
Probabilistic models satisfying this postulate 
are sometimes called \emph{sharp} \cite{Wilce2019conjugatesfilters}.
\vspace{0,3 cm}

\noindent
\textbf{Axiom 1:} 
\itshape
For each minimal extreme point 
$e \in ext(\left[0,\mathbb{I}\right])$, 
there is only one single state $\mathbb{P}_e$ 
with $\mathbb{P}_e (e) = 1$.
\normalfont
\vspace{0,3 cm}

\noindent
\textbf{Lemma 2:} 
\itshape\itshape
$\mathbb{P}_e$ becomes a pure state 
for each minimal extreme point $e$ in $ext(\left[0,\mathbb{I}\right])$.
\vspace{0,3 cm}
\normalfont

Proof. Suppose $e$ is a minimal extreme point $e$ in $ext(\left[0,\mathbb{I}\right])$
and \linebreak
$\mathbb{P}_e = s \mu_1 + (1-s) \mu_2$
with $\mu_1, \mu_2 \in S_A$, $0 < s < 1$.
Then 
$ 1 = \mathbb{P}_e (e) = s \mu_1 (e) + (1-s) \mu_2 (e)$
and, since $0 \leq \mu_{1,2}(e) \leq 1$, we get 
$1 = \mu_1(e) = \mu_2(e)$. Axiom 1 implies $\mu_1 = \mathbb{P}_e = \mu_2$.
Therefore $\mathbb{P}_e$ is an extreme point of $S_A$. \hfill $\square$
\vspace{0,3 cm}

We now come to our second axiom which just means that each pure state 
has the form $\mathbb{P}_e$ with some 
minimal extreme point $e \in ext(\left[0,\mathbb{I}\right])$.
\vspace{0,3 cm}

\noindent
\textbf{Axiom 2:} 
\itshape
For each pure state $\mu \in S_A$, there is a minimal extreme point $e$ 
in~$ext(\left[0,\mathbb{I}\right])$ such that $\mathbb{P}_e = \mu$.
\vspace{0,3 cm}
\normalfont

An important consequence of Axiom 2 is that $ext(\left[0,\mathbb{I}\right])$ 
is rather large. Without Axiom 2, there could be no more 
than the two extreme points $0$ and $\mathbb{I}$.
However, by the Krein-Milman theorem \cite{AS01}, 
the state space $S_A$ contains many extreme points,
and then by Axiom 2, $ext(\left[0,\mathbb{I}\right])$ 
includes many (minimal) extreme points. 

For order unit spaces $A$ that satisfy the Axioms 1 and 2, 
we now introduce the following property which will play an important role in this paper.
\newpage
\itshape
\begin{enumerate}
	\item [($\ast$)] If $\mathbb{P}_e (a) = 1$ holds for $a \in \left[0,\mathbb{I}\right]$
and a minimal extreme point $e \in ext(\left[0,\mathbb{I}\right])$, 
then $e \leq a$.
\end{enumerate}
\normalfont
\noindent
Some first consequences of ($\ast$) are presented in the next lemma.
\vspace{0,3 cm}

\noindent
\textbf{Lemma 3:} 
\itshape
Let $A$ be an order unit space $A$ that satisfies 
Axiom 1, Axiom 2 and possesses the property ($\ast$).
\begin{enumerate}
	\item [(i)] For any minimal extreme points $e_1, e_2 \in ext(\left[0,\mathbb{I}\right])$ 
we have: 
\begin{center}
$\mathbb{P}_{e_1}(e_2) = 1 \Leftrightarrow e_1 = e_2$, and
\end{center}
\begin{center}
$\mathbb{P}_{e_1}(e_2) = 0 \Leftrightarrow \mathbb{P}_{e_2}(e_1) = 0 \Leftrightarrow e_1 + e_2 \leq \mathbb{I}$.
\end{center}
	\item [(ii)] 
Suppose $q,e \in ext(\left[0,\mathbb{I}\right])$ with a minimal $e$. 
\begin{center}
If $q + e \leq \mathbb{I}$, then $q + e \in ext(\left[0,\mathbb{I}\right])$.
\end{center}
\begin{center}
If $e \leq q$, then $q - e \in ext(\left[0,\mathbb{I}\right])$.
\end{center}
	\item [(iii)] If $e_1, ..., e_n \in ext(\left[0,\mathbb{I}\right])$ are minimal
and $\sum^{n}_{k=1} e_k \leq \mathbb{I}$, then
$\sum^{n}_{k=1} e_k \in ext(\left[0,\mathbb{I}\right])$.
	\item [(iv)] For each $a \in A$  there is a minimal extreme point 
$e \in ext(\left[0,\mathbb{I}\right])$ 
with $\left|\mathbb{P}_e (a)\right| = \left\|a\right\| $.
For $0 \leq a$ we have $ \mathbb{P}_e (a) = \left\|a\right\| $
and $ \left\|a\right\| e \leq a$.
\end{enumerate}
\normalfont
Proof. 
(i) The first part immediately follows from ($\ast$) and the minimality of $e_2$. 
Now suppose
$\mathbb{P}_{e_1}(e_2) = 0$ for the minimal extreme points 
$e_1, e_2 \in ext(\left[0,\mathbb{I}\right])$. 
Then 
$ \mathbb{P}_{e_1}( \mathbb{I}- e_2) = 1$ and 
$e_1 \leq \mathbb{I}- e_2$ by ($\ast$).
This means $e_1 + e_2 \leq \mathbb{I}$. Furthermore,
$1 = \mathbb{P}_{e_2}(e_1) + \mathbb{P}_{e_2}(e_2) = \mathbb{P}_{e_2}(e_1) + 1$
and $ \mathbb{P}_{e_2}(e_1) = 0 $.
With exchanged roles of $e_1$ and $e_2$, we get $\mathbb{P}_{e_1}(e_2) = 0$
from $ \mathbb{P}_{e_2}(e_1) = 0 $.

(ii) Suppose $q,e \in ext(\left[0,\mathbb{I}\right])$, $e$ minimal 
and $q + e \leq \mathbb{I}$. Then \linebreak
$1 \geq \mathbb{P}_{e}(q) + \mathbb{P}_{e}(e) = \mathbb{P}_{e}(q) + 1$
and thus $\mathbb{P}_{e}(q)= 0$. 
Now assume 
$q + e = sa + (1-s) b$ with $0 < s < 1$ and $a,b \in \left[0,\mathbb{I}\right]$.
Then 
$1 =\mathbb{P}_{e}(q + e) =  s \mathbb{P}_{e}(a) + (1-s) \mathbb{P}_{e}(b)$
and from $\mathbb{P}_{e}(a), \mathbb{P}_{e}(b) \leq 1$ we get
$\mathbb{P}_{e}(a) = \mathbb{P}_{e}(b) = 1$.
The property ($\ast$) implies $e \leq a$ and $e \leq b$.
Then we have $q= s(a-e) + (1-s)(b-e)$ with $a-e, b-e \in \left[0,\mathbb{I}\right]$
for the extreme point $q$. Therefore $q = a - e = b - e$
and $q + e = a = b$ and we have shown that $q + e$ is an extreme point.

Now suppose $q,e \in ext(\left[0,\mathbb{I}\right])$, $e$ minimal, 
$e \leq q $ and $q - e = sa + (1-s) b$ with $0 < s < 1$ 
and $a,b \in \left[0,\mathbb{I}\right]$. 
Then $\mathbb{P}_e(q) = 1$ and
$0 = \mathbb{P}_e(q -e) = s \mathbb{P}_e(a) + (1-s) \mathbb{P}_e(b)$.
Thus $0 =\mathbb{P}_e(a) = \mathbb{P}_e(b)$ and 
$1 =\mathbb{P}_e(\mathbb{I} - a) = \mathbb{P}_e(\mathbb{I} -b)$.
From ($\ast$) we get $e \leq \mathbb{I} -a$ and $e \leq \mathbb{I} -b$
or, equivalently, $a + e \leq \mathbb{I} $ and $ b + e \leq \mathbb{I}$. 
Since $q$ is an extreme point, we get from $q = s(a+e) + (1-s)(b+e)$ 
that $q = a+e = b+e $ must hold. Therefore we have
$q - e = a = b$ and we have shown that $q$ is an extreme point.

(iii) follows from (ii) by complete induction.

(iv) Let $a \in A$. As in any order unit space there is a state $\mu$
with $ \left|\mu(a)\right| = \left\|a\right\| $. 
Then either $ \mu(a) = \left\|a\right\| $ or $ \mu(a) = - \left\|a\right\| $.

In the first case, we consider set $\left\{\mu \in S_A : \mu(a) = \left\| a \right\| \right\} \subseteq S_A$,
which is non-empty, convex and compact. 
By the Krein-Milman theorem \cite{AS01},  
it contains at least one extreme point $\mu_0$.
We now show that $\mu_0$ is an extreme point of $S_A$ and assume that
$\mu_0 = s \mu_1 + (1-s) \mu_2$ with $\mu_1, \mu_2 \in S_A$ and $0 < s < 1$.
Then 
$\left\|a\right\| = \mu_0(a) = s \mu_1(a) + (1-s) \mu_2(a)$.
Since $\mu_1(a) \leq \left\|a\right\| $ and $\mu_2(a) \leq \left\|a\right\| $, 
we get $\mu_1(a) = \left\|a\right\| = \mu_2(a)$
and $\mu_1, \mu_2$ lie in the subset of $S_A$ where $\mu_0$ is an extreme point.
Therefore, $\mu_1 = \mu_2 = \mu_0$ and $\mu_0$ becomes a pure state.
By Axiom 2, there is a minimal extreme point $e \in ext(\left[0,\mathbb{I}\right])$ 
with $\mathbb{P}_e = \mu_0$. Then 
$\left\|a\right\| = \mu_0(a) = \mathbb{P}_e(a)$.

In the second case, we consider set 
$\left\{\mu \in S_A : \mu(a) = - \left\| a \right\| \right\} $
and proceed in the same way as above.
Finally, the property ($\ast$) 
implies $ \left\|a\right\| e \leq a$. \hfill $\square$
\vspace{0,3 cm}

The finite dimensional JBW algebras coincide with the
formally real (also named Euclidean) 
Jordan algebras \cite{AS02, hanche1984jordan}
and form order unit spaces that satisfy the axioms 1 and 2. 
The extreme points of their 
unit intervals are the idempotent elements.
In Ref. \cite{nie2021generic} it was shown that the identities 
$ \mathbb{P}_e (a) = s$ and $\left\{e,a,e\right\} = se$ 
are then equivalent for any minimal idempotent element $e$, 
any algebra element $a$ and $s \in \mathbb{R}$.
Here, $\left\{\ ,\ ,\ \right\}$ denotes 
the so-called triple product in the Jordan algebra:
$\left\{a,b,c\right\} := a \circ (b \circ c) - b \circ (c \circ a) + c \circ (a \circ b) $ 
for the algebra elements $a,b,c$.
When the Jordan product $\circ$ stems from an associative product via
$a \circ b := (ab+ba)/2$ (this means that the Jordan algebra is \emph{special}),
the triple product $\left\{a,b,a\right\}$ coincides 
with the simple operator product $aba$.
This does not hold in the \emph{exceptional} Jordan algebra 
formed by the Hermitian (self-adjoint) 
$3\times3$-matrices over the octonions.
The property ($\ast$) follows from 1.38 Proposition (identity 1.49) in Ref. \cite{AS02}:
$ 0 \leq a \leq \mathbb{I}$ and $\mathbb{P}_e (a) = 1$.
Then $\mathbb{P}_e (\mathbb{I} - a) = 0$ and $\left\{e,\mathbb{I}-a,e\right\} = 0$.
Now use 1.49 from \cite{AS02} to get
$\mathbb{I}-a = \left\{e',\mathbb{I}-a,e'\right\} \leq e' = \mathbb{I}-e$ 
and thus $e \leq a$.

\section{Compact convex sets}

Let $\Omega$ be any compact convex set in some locally convex real vector space $V$.
By the Krein-Milman theorem \cite{AS01}, 
$\Omega$ is the closed convex hull of its extreme points
and thus contains at least one extreme point unless $\Omega = \emptyset$.
Here $A_\Omega$ shall denote the order unit space of all 
\emph{continuous affine} functions on $\Omega$; its order unit
is the constant function $\mathbb{I} \equiv 1$.
For each $\omega \in \Omega$ we define the following function $e_\omega$ on $\Omega$:
$$e_\omega (\zeta) := inf\left\{a(\zeta) : a \in A_\Omega, 
0 \leq a \text{ and } a(\omega)=1 \right\}$$
for $\zeta \in \Omega$. 
Since $\mathbb{I}(\omega) = 1$,
we then have $e_\omega (\zeta) \leq 1$ for all $\zeta \in \Omega$.
Generally, this function is neither continuous nor affine
and does not belong to $A_\Omega$. We now introduce the following novel property of a 
compact convex set $\Omega$:
\itshape
\begin{enumerate}
	\item [($\ast \ast$)] For each extreme point $\omega \in \Omega$, 
the function $e_\omega$ is contained in $A_\Omega$ 
and $e_\omega(\zeta) \neq 1$ for all $\zeta \in \Omega$ with $\zeta \neq\omega$.
\end{enumerate}
\normalfont
This is a technical mathematical condition, 
the relevance of which is revealed in the following theorem. 
Some better motivated conditions that are equivalent to it 
in certain situations are presented in section 5.
\vspace{0,3 cm}

\noindent
\textbf{Theorem 1:} 
\itshape
\begin{enumerate}
	\item [(i)] Let $A$ be an order unit space that satisfies Axiom 1, Axiom 2 
and possesses the property ($\ast$). The state space $S_A$ then possesses
the property ($\ast \ast$).
	\item [(ii)] Let $\Omega$ be a compact convex set (in some locally convex space) 
with the property ($\ast \ast$). 
The order unit space $A_\Omega$ then satisfies Axiom 1, Axiom 2 
and possesses the property ($\ast$). Moreover, the state space of $A_\Omega$ 
is isomorphic to $\Omega$.
\end{enumerate}
\normalfont
Proof. (i) Let $A$ be as in part (i) of the theorem. 
For each $a \in A$ define the function $\hat{a}$ on $S_A$ 
by $\hat{a}(\rho) := \rho(a)$, $\rho \in S_A$.
The map $A \ni a \rightarrow \hat{a} \in A_{S_A}$
is an isomorphism of the order unit spaces $A$ 
onto a dense linear subspace of $A_{S_A}$ (Theorem 1.20 in \cite{AS01}).

Now let $\mu$ be an extreme point of the state space $S_A$.
By Axiom 2 there is a minimal $q \in ext(\left[0,\mathbb{I}\right])$ 
with $\mu = \mathbb{P}_q$
and then consider $\hat{q}$. For $a \in A$ with 
$0 \leq a \leq \mathbb{I}$ and $1 = \hat{a}(\mu) = \mu(a) = \mathbb{P}_q(a)$
we get $q \leq a$ by ($\ast$) and thus $\hat{q} \leq \hat{a}$. 
Then $\hat{q} \leq b$ for all $b \in A_{S_A}$ with 
$0 \leq b $ and $b(\mu) = 1$,
because $\hat{A}$ is dense in $A_{S_A}$.
Since $\hat{q}(\mu) = \mu(q) = \mathbb{P}_q (q) =1$, this means that
$\hat{q}$ is the infimum $e_\mu$ and therefore we have $e_\mu \in A_{S_A}$.
Moreover $e_\mu (\rho) = \hat{q}(\rho) =\rho(q) \neq 1$ 
for the states $\rho \neq \mu = \mathbb{P}_q$ by Axiom 1.

(ii) Let $\Omega$ be a compact convex set with the property ($\ast \ast$).
First we show that $e_\omega$ with $\omega \in ext(\Omega)$ is an 
extreme point of the unit interval in $A_\Omega$.
Suppose $\omega \in ext(\Omega)$ and $e_\omega = sa + (1-s)b$ 
with $a,b \in A_\Omega$, $0 \leq a,b \leq \mathbb{I}$, $0 < s < 1$.
Then $a(\omega)=1=b(\omega)$ since $e_\omega(\omega) = 1$,
and from ($\ast \ast$) we get $e_\omega(\zeta) \leq a(\zeta)$ and
 $e_\omega(\zeta) \leq b(\zeta)$ for all $\zeta \in \Omega$.
From $e_\omega(\zeta) = sa(\zeta) + (1-s)b(\zeta)$ we then get
$e_\omega(\zeta) = a(\zeta)$ and $e_\omega(\zeta) = b(\zeta)$ 
for each $\zeta \in \Omega$. This means $a = b = e_\omega$ and 
thus $e_\omega$ is an extreme point.

For $\omega \in ext(\Omega)$ we now show that $e_\omega$ is a minimal 
extreme point in $\left[0,\mathbb{I}\right] \subseteq A_\Omega$. 
Suppose $0 \neq q \leq e_\omega$ with $q \in ext(\left[0,\mathbb{I}\right])$. 
If $q(\omega) = 1$, we get from ($\ast \ast$) that $e_\omega \leq q$ and 
thus $q = e_\omega$. If $q(\omega) < 1$, we have from ($\ast \ast$) that
$q(\zeta) \leq e_\omega(\zeta) < 1$ for all $\omega \neq \zeta \in \Omega$
and thus $q(\zeta) < 1$ for all $\zeta \in \Omega$. 
Then $\left\|q\right\| = sup \left\{\left|q(\zeta)\right| : \zeta \in \Omega \right\} < 1$
and $q \notin ext(\left[0,\mathbb{I}\right])$ or $q = 0$.
Note that, owing to the continuity of $q$ and the compactness of $\Omega$,
the function $q$ assumes its maximum at some point in $\Omega$.

Vice versa, each minimal extreme point
of the unit interval in $A_\Omega$ has the form $e_\omega$ 
with $\omega \in ext(\Omega)$. To prove this, 
let $p$ be such an minimal extreme point.
The affine and continuous function $p$ assumes its maximum in an 
extreme point $\omega \in \Omega$. 
Since $0 \neq p \in ext(\left[0,\mathbb{I}\right])$, this maximum is $1$
and thus $e_\omega \leq p$ by ($\ast \ast$). Since $p$ is minimal, 
we get $e_\omega = p$.

For $\omega \in \Omega$ we define $\delta_\omega(a) := a(\omega)$
for every $a \in A_\Omega$. 
Note that \linebreak
$\delta_{s \omega_1 + (1-s) \omega_2}(a) 
= a(s \omega_1 + (1-s) \omega_2) 
= s a(\omega_1) + (1-s) a(\omega_2)
= s \delta_{\omega_1}(a) + (1-s) \delta_{\omega_2}(a)$
for $\omega_1, \omega_2 \in \Omega$ and $0 \leq s \leq 1$
and that
$\delta_{\omega_\alpha}(a) = a(\omega_\alpha) \rightarrow a(\omega_o) = \delta_{\omega_o}(a)$
for $\omega_\alpha \rightarrow \omega_o$.
Moreover, every state on $A_\Omega$ can be represented as $\delta_\omega$ 
with some $\omega \in \Omega$ (Lemma 8.70 in \cite{AS02}).
Therefore the map $\Omega \ni \omega \rightarrow \delta_\omega$
is a continuous affine isomorphism between $\Omega$ and the state space of $A_\Omega$
and allocates the extreme points of $\Omega$ to the pure states of $A_\Omega$.

Axiom 1 is satisfied because,
with $\omega \in ext(\Omega)$, $\delta_\omega$ is the only state 
allocating the numerical value $1$ to $e_\omega$ by ($\ast \ast$).
Axiom 2 holds with $e_\omega$ for the pure state $\delta_\omega$, $\omega \in ext(\Omega)$.
The property ($\ast$) is a direct consequence of ($\ast \ast$)
since $\mathbb{P}_{e_\omega} = \delta_\omega$ 
for $\omega \in ext(\Omega)$. \hfill $\square$
\vspace{0,3 cm}

\begin{figure}[ht]
\centering
\begin{minipage}[t]{0.30\linewidth}
\centering
\includegraphics[width=2.3cm]{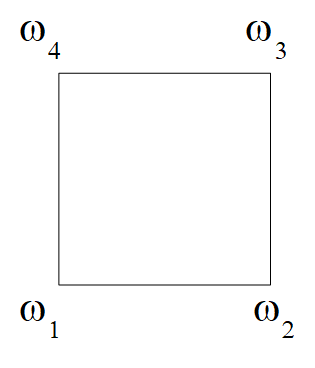}
\caption{Square}
\end{minipage}
\hfill
\begin{minipage}[t]{0.30\linewidth}
\centering
\includegraphics[width=3.0cm]{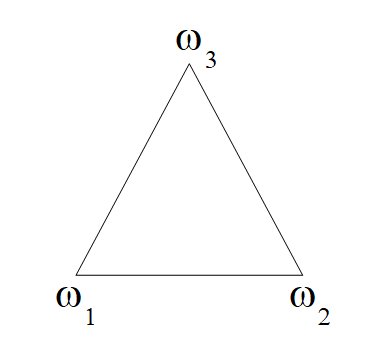}
\caption{Triangle}
\end{minipage}
\hfill
\begin{minipage}[t]{0.30\linewidth}
\centering
\includegraphics[width=3.0 cm]{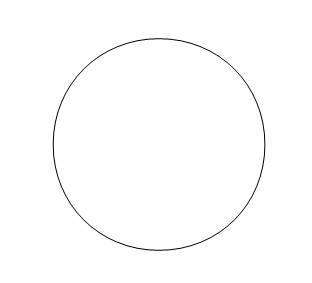}
\caption{Disk}
\end{minipage}
\end{figure}

To get a better understanding of the property ($\ast \ast$),
let us have a look at the three simple examples shown in the Figures 1, 2 and 3.

The extreme points of the square (Fig. 1) are 
$ \omega_1 , \omega_2 , \omega_3 , \omega_4 $.
A first affine function on the square 
with values in the interval $\left[0,1\right] \subseteq \mathbb{R}$
is the one allocating $1$ to both $\omega_1$ and $\omega_2$ and 
$0$ to both $\omega_3$ and $\omega_4$, a second one 
allocates $1$ to both $\omega_1$ and $\omega_4$
and $0$ to both $\omega_2$ and $\omega_3$.
Since $e_{\omega_1}$ lies below these two,
we have $ 0 = e_{\omega_1}(\omega_2) = e_{\omega_1}(\omega_2) =e_{\omega_1}(\omega_3)$, 
but $e_{\omega_1}$ cannot be affine then. Therefore, 
the square does not possess the property ($\ast \ast$).
Although the square does not fulfill our requirements, 
it is sometimes considered the state space of a 
generalized bit~\cite{barrett2007information}.

The situation is different with the triangle (Fig. 2).
Its extreme points are $ \omega_1 , \omega_2 , \omega_3 $.
For $k=1,2,3$, $e_{\omega_k}$ is the affine function 
with $e_{\omega_k}(k)=1$ and $e_{\omega_k}(k')=0$ for $k'\neq k$
and the triangle does possess the property ($\ast \ast$).

The triangle is the $3$-\emph{simplex} 
and represents a \emph{classical} finite dimensional state space 
like any other $n$-simplex ($n \in \mathbb{N}$). 
The associated order unit space 
[ $\mathbb{R}^{n}$ with the cone 
$\left\{(r_1,...,r_n) : 0 \leq r_k \text{ for } k=1,...,n \right\}$
and the order unit $(1,...,1)$ ] 
possesses the property ($\ast$)
and the $n$-simplex has the property~($\ast \ast$).

Each point on the boundary of the disk (Fig. 3) 
is an extreme point. For any point $\omega$,
$e_\omega$ is the affine function with values 
in the interval $\left[0,1\right]$
that allocates $1$ to $\omega$ and $0$ to its antipodal point.
Thus the disk does possess the property~($\ast \ast$).

The disk is just one example in the class of the strictly convex and smooth
compact convex sets \cite{AS02}. Here the extreme boundary 
coincides with the topological boundary and $e_\omega$ is the continuous affine
function allocating $1$ to the extreme point $\omega$ and 
$0$ to its antipodal point. This class gives rise to the
generalized qubit models studied in Ref. \cite{Nie2022genqubit}.
We shall come back to this at the end of section 5.

Although ($\ast \ast$) looks like a rather simple geometric property,
it turns out that it is very hard to verify for a given compact convex set
that is not strictly convex and smooth.
From Theorem 1 and the remarks at the end of section 2 we now know
that the state spaces of the formally real Jordan algebras 
with finite dimension possess this property.
The state spaces of the associative Jordan algebras are the simplexes.

Already fifty years ago, Mielnik introduced the function $e_\omega$ 
with an extreme point $\omega$ in any compact convex set and interpreted 
$e_{\omega_1}(e_{\omega_2})$ with two extreme points $\omega_1$ and $\omega_2$ 
as a generalized quantum transition probability 
\cite{mielnik1969theory, mielnik1974generalized}, 
but he did not require that $e_\omega$ is affine (property ($\ast \ast$)
of the convex set).

\section{Finite information capacity}

A finite subset $ \left\{p_1, p_2, ..., p_n\right\} \subseteq ext(\left[0,\mathbb{I}\right])$
in an order unit space $A$ shall be called \emph{orthogonal},
if $ \sum^{n}_{k=1} p_k \leq \mathbb{I}$.
Any family of elements in $ext(\left[0,\mathbb{I}\right])$
shall be called orthogonal, if each finite subfamily is orthogonal.
\vspace{0,3 cm}

\noindent
\textbf{Lemma 4:} 
\itshape
Let $A$ be an order unit space 
that satisfies Axiom 1 and Axiom~2.
Suppose $A \ni a = \sum^{n}_{k=1} s_k e_k$ with
$s_k \in \mathbb{R}$, 
minimal extreme points $e_k \in ext(\left[0,\mathbb{I}\right])$,
$k = 1,2,...,n$ ($n \in \mathbb{N}$) and $\sum^{n}_{k=1} e_k \leq \mathbb{I}$. Then:
\begin{enumerate}
	\item [(i)] $\mathbb{P}_{e_k}(a) = s_k$ for $k = 1,...,n$.
	\item [(ii)] $\left\| a \right\| = max \left\{ \left| s_k \right| : k=1,...,n \right\}$.
	\item [(iii)] $0 \leq a$ iff $0 \leq s_k$ for $k = 1,...,n$.
\end{enumerate}
\normalfont
Proof.
(i) This follows immediately from Lemma 3 (i).

\noindent
(ii) Suppose $ \left|s_j\right| = max \left\{ \left| s_k \right| : k=1,...,n \right\}$.
Then $a \leq \left|s_j\right| \sum^{n}_{k=1} e_k \leq \left|s_j\right| \mathbb{I}$
and $a \geq - \left|s_j\right| \sum^{n}_{k=1} e_k \geq - \left|s_j\right| \mathbb{I}$.
Therefore $ \left\|a\right\| \leq \left|s_j\right| $. 
Since $\mathbb{P}_{e_j}$ is a state with $\mathbb{P}_{e_j}(a) = s_j$, 
we have $ \left\|a\right\| = \left|s_j\right| $.

\noindent
(iii) If $0 \leq a$, then $0 \leq \mathbb{P}_{e_k}(a) = s_k$ for $k = 1,...,n$.
If $0 \leq s_k$ for $k = 1,...,n$, 
then $a = \sum^{n}_{k=1} s_k e_k \geq 0$. \hfill $\square$
\vspace{0,3 cm}

\noindent
\textbf{Lemma 5:}
\itshape
Suppose the order unit space
$A$ satisfies Axiom 1, Axiom 2 and ($\ast$).
A family of minimal elements in $ext(\left[0,\mathbb{I}\right])$
becomes orthogonal iff each pair in this family is orthogonal. 
\vspace{0,3 cm}
\normalfont

Proof.
Suppose $e_1, e_2, ...$ are pairwise orthogonal 
and minimal in $ext(\left[0,\mathbb{I}\right])$. 
Then $ e_1 + e_2 \leq \mathbb{I} $. 
We now proceed with complete induction.
Assume 
$\sum^{n-1}_{k=1} e_{k} \leq  \mathbb{I} $.
By Lemma 3 (i), $\mathbb{P}_{e_{k'}}(e_{k}) = 0$ for $k' \neq k$
and then 
$\mathbb{P}_{e_n}(\mathbb{I} - \sum^{n-1}_{k=1} e_{k}) = 1$.
Finally by ($\ast$) 
$e_n \leq \mathbb{I} - \sum^{n-1}_{k=1} e_{k}$.
\hfill $\square$
\vspace{0,3 cm}

Following the notation of Ref. \cite{Barnum2020composites},
the maximum cardinality of the orthogonal families 
of non-zero elements in $ext(\left[0,\mathbb{I}\right])$
shall here be called the \emph{information capacity} of $A$.
Note that each orthogonal family is linearly independent. 
If $$s_1 p_1 + s_2 p_2 + ... + s_n p_n = 0$$ for the
orthogonal elements $p_1,p_2,...,p_n \in ext(\left[0,\mathbb{I}\right])$ 
and $s_1,s_2,...,s_n \in \mathbb{R}$, use states $\mu_k$ with $\mu_k(p_k)=1$.
From $$1 \geq \sum^{n}_{k=1} \mu_{k'}(p_k) = 1 + \sum_{k \neq k'} \mu_{k'}(p_k)$$
we get $\mu_{k'}(p_{k})=0$ for $k \neq k'$ and therefore $s_k = 0$ for each $k$.
This means that the information capacity of $A$ is finite, 
if $A$ has a finite dimension.

However, the dimension of $A$ need not be finite, 
if the information capacity is finite.
Dimension and information capacity are completely different things, 
where the information capacity 
cannot exceed the dimension. The so-called spin factors  
form the JBW algebras with information capacity two 
and can have any dimension of any cardinality \cite{AS02, hanche1984jordan}.

As usual, for two or more elements, we denote by $\vee$ 
the supremum (or least upper bound) in $ext(\left[0,\mathbb{I}\right])$)
and by $\wedge$ the infimum (or greatest lower bound)
in~$ext(\left[0,\mathbb{I}\right])$), provided that it exists.
\vspace{0,3 cm}

\noindent
\textbf{Lemma 6:}
\textit{Let $A$ be an order unit space 
with finite information capacity $m$ which
satisfies Axiom 1 and Axiom 2 
and possesses the property ($\ast$).
\begin{itemize}
	\item [(i)]
Each $p \in ext(\left[0,\mathbb{I}\right])$ with $p \neq 0$ 
has the form $p = \sum^{n}_{k=1} e_k$ 
with minimal extreme points $e_k$ in~$\left[0,\mathbb{I}\right]$
and $n \leq m$.
	\item [(ii)]
If $p_j \in ext(\left[0,\mathbb{I}\right])$, $j = 1,2,...,n$,
are pairwise orthogonal, they form an orthogonal family, 
$\sum^{n}_{j=1} p_j \in ext(\left[0,\mathbb{I}\right])$
and $ \sum^{n}_{j=1} p_j = \vee^{n}_{j=1} p_j$.
	\item [(iii)]
If $p,q \in ext(\left[0,\mathbb{I}\right])$ with $q \leq p$,
then $p - q \in ext(\left[0,\mathbb{I}\right])$ and $p - q = p \wedge (\mathbb{I}-q)$.
\end{itemize}}

Proof. 
(i) Suppose $0 \neq p \in ext(\left[0,\mathbb{I}\right])$.
By Lemma 3 (iv) there is a minimal $e_1 \in ext(\left[0,\mathbb{I}\right])$ 
with $e_1 \leq p$. If $p = e_1$, we already have the desired form.
If $p \neq e_1$, we have $p - e_1 \in ext(\left[0,\mathbb{I}\right])$ 
from Lemma 3 (ii) and we get a further minimal 
$e_2 \in ext(\left[0,\mathbb{I}\right])$ 
with $e_2 \leq p - e_1$. If $e_2 = p - e_1$, we are done.
If not, we continue this process, which must stop at latest
after we have found $m$ orthogonal minimal elements 
$e_1, e_2, ... \in ext(\left[0,\mathbb{I}\right])$.

(ii) Suppose $p_j \in ext(\left[0,\mathbb{I}\right])$, $j = 1,2,...,n$,
are pairwise orthogonal. By (i) each $p_j \neq 0$ is a sum of 
minimal extreme points in $\left[0,\mathbb{I}\right]$.
These are pairwise orthogonal, even if they belong to different $p_j$,
since the $p_j$ are pairwise orthogonal.
By Lemma 5, the whole family becomes orthogonal,
and from Lemma 3 (iii) we get 
$\sum^{n}_{j=1} p_j \in ext(\left[0,\mathbb{I}\right])$.

Now assume $p_j \leq q$ for $j = 1,2,...,n$ 
with $q \in ext(\left[0,\mathbb{I}\right])$.
Then the set $\left\{\mathbb{I}-q, p_1, p_2,.... p_n\right\}$ 
is pairwise orthogonal and becomes an orthogonal family 
from the above arguments. Therefore 
$\mathbb{I}-q + \sum^{n}_{j=1} p_j \leq \mathbb{I}$ and $\sum^{n}_{j=1} p_j \leq q$.
Thus we have shown that the sum is the least upper bound.

(iii) Suppose $p,q \in ext(\left[0,\mathbb{I}\right])$ with $q \leq p$.
By (i) we have $q = \sum^{n}_{k=1} e_k$ 
with minimal extreme points $e_k$ in $\left[0,\mathbb{I}\right]$.
From Lemma 3 (ii) we get by complete induction
$p - q = p - \sum^{n}_{k=1} e_k \in ext(\left[0,\mathbb{I}\right])$.
Moreover, $ p - q = \mathbb{I} - ((\mathbb{I} - p) \vee q) = p \wedge (\mathbb{I}-q) $.
\hfill $\square$

\section{Spectrality}

Lemma 6 (iii) implies that
$ext(\left[0,\mathbb{I}\right])$ is an 
\emph{orthomodular} partially ordered set~\cite{foulis1992filters}
with the \emph{orthocomplementation} $'$, defined by $ p' := \mathbb{I} - p $.
Thus $ext(\left[0,\mathbb{I}\right])$ is an excellent candidate 
for a so-called \emph{quantum logic}.
In this section we shall see that $ext(\left[0,\mathbb{I}\right])$ 
actually becomes a \emph{lattice} \cite{AS02},
which means that $p \vee q$ always exists 
for $p,q \in ext(\left[0,\mathbb{I}\right])$ 
(not only when $p$ and $q$ are orthogonal).

The minimal elements in a quantum logic are usually 
called \emph{atoms} \cite{AS02, piron1964axiomatique}
and a quantum logic is said to be \emph{atomic}, if an atom lies under
each non-zero element in the quantum logic. In our case, the atoms
become identical with the minimal extreme points 
of $\left[0,\mathbb{I}\right]$.

An order unit space $A$ with finite information capacity
shall here be called \emph{spectral},
if each $a \in A$ can be represented as 
$$ a = \sum^{n}_{k=1} s_k e_k $$ with
$s_k \in \mathbb{R}$, 
minimal extreme points $e_k \in ext(\left[0,\mathbb{I}\right])$,
$k = 1,2,...,n$, $n \in \mathbb{N}$, and $\sum^{n}_{k=1} e_k \leq \mathbb{I}$. 
The coefficicients $s_k$ represent the potential numerical
measurement outcomes. The probability for the outcome $s_k$
in a given state $\mu$ is $\sum_{j:s_j=s_k} \mu(e_j)$.

Spectrality is a desirable property, 
since it ensures that, with a given state, each element of $A$ 
possesses a probability distribution over 
some potential numerical measurement outcomes
and can thus be interpreted as a physical observable. 
The above spectral representation is unique only if
$s_k \neq s_{k'}$ for $k \neq k'$. In order to achieve uniqueness
in the other cases, one must consider another representation 
with the combined summands $s_k \sum_{j:s_j=s_k} e_j$,
where the $e_j$ in each sum $\sum_{j:s_j=s_k} e_j$
are not uniquely determined. This is the same situation as 
with the observables with discrete spectra 
in ordinary quantum mechanics.

Following the notation in Refs. \cite{nie2021generic, Nie2022genqubit},
the state space $S_A$ of an order unit space $A$ is called \emph{strong} if 
$$ \left\{\mu \in S_A : \mu(p) = 1 \right\} \subseteq \left\{\mu \in S_A : \mu(q) 
= 1 \right\} \ \Rightarrow \ p \leq q$$
holds for $p,q \in ext(\left[0,\mathbb{I}\right])$.
\vspace{0,3 cm}

\noindent
\textbf{Theorem 2:} 
\itshape
Let $A$ be an order unit space 
with finite information capacity $m$ and suppose 
$A$ satisfies Axiom 1 and Axiom 2. 
\begin{enumerate}
	\item [(i)] $A$ possesses the property ($\ast$) iff
$A$ is spectral and the state space $S_A$ is strong. 
	\item [(ii)] If one of the equivalent conditions of (i) holds,
$ext(\left[0,\mathbb{I}\right])$ is an atomic orthomodular lattice.
\end{enumerate}
\normalfont
\vspace{0,3 cm}

Proof. 
(i) Suppose $A$ has the property ($\ast$).
First we assume $0 \leq a$ for $a \in A$.
Define $s_1 := \left\|a\right\| $.
By Lemma~3~(iv) there is a minimal extreme point 
$e_1 \in ext(\left[0,\mathbb{I}\right])$ 
with $\mathbb{P}_{e_1}(a) = s_1$ and $s_1 e_1 \leq a$.
Then we consider $a_1 := a - s_1 e_1 \geq 0$.
If $a_1 = 0$, we have already found the spectral form of $a$.
Thus we can assume $a_1 \neq 0$ and,
with $s_2 := \left\|a_1\right\| $, there is a further
minimal extreme point 
$e_2 \in ext(\left[0,\mathbb{I}\right])$ 
with $\mathbb{P}_{e_2}(a_1) = s_2$ and $s_2 e_2 \leq a_1$.
From $\mathbb{P}_{e_1}(a_1) = 0$ and $s_2 \neq 0$ we get 
$\mathbb{P}_{e_1}(e_2) = 0$.
Then we consider $a_2 := a_1 - s_2 e_2 = a - s_1 e_1 - s_2 e_2 \geq 0$.
If $a_2 = 0$ our job is done and we can assume $a_2 \neq 0$.
With $s_3 := \left\|a_2\right\| $
there is again a minimal extreme point 
$e_3 \in ext(\left[0,\mathbb{I}\right])$ with 
$\mathbb{P}_{e_3}(a_2) = s_3$ and
$s_3 e_3 \leq a_2$.
From $\mathbb{P}_{e_k}(a_2) = 0$ and $s_3 \neq 0$
we get $\mathbb{P}_{e_k}(e_3) = 0$ for $k = 1,2$. 
By Lemma 3 (i) $e_1, e_2, e_3$ are orthogonal.
If $a_3 = 0$, we are done. In the other case,
we continue this process until $a_{n} = 0$.
This happens at latest when $n=m+1$.

For any $a \in A$ we have $0 \leq a + \left\|a\right\| \mathbb{I}$ and
$a + \left\|a\right\| \mathbb{I} = \sum^{n}_{k=1} s_k e_k$. 
Then
$a = \sum^{n}_{k=1} (s_k - \left\|a\right\| ) e_k - \left\|a\right\| (\mathbb{I} - \sum^{n}_{k=1} e_k) $.
Since $\mathbb{I} - \sum^{n}_{k=1} e_k \in ext(\left[0,\mathbb{I}\right])$
by Lemma 1 and Lemma 3 (iii), we can use Lemma 6 (i)
to find further $e_k$, $k = n+1, ..., l$, with 
$\mathbb{I} - \sum^{n}_{k=1} e_k = \sum^{l}_{k=n+1} e_k$.
Then all $e_k$, $k=1,...,l$, are orthogonal 
and we have found the spectral form of~$a$. 

To prove that $S_A$ is strong, assume that
\begin{center}
$ \left\{\mu \in S_A : \mu(q) = 1 \right\}  \subseteq \left\{\mu \in S_A : \mu(p) = 1 \right\}$
\end{center}
holds for $p,q \in ext(\left[0,\mathbb{I}\right])$. 
We have to show that $q \leq p$.
The case $q = 0$ is trivial and we can assume $q \neq 0$.
By Lemma~6~(i), $q$ has the form
$q = \sum^{n}_{k=1} e_k$ 
with minimal extreme points $e_k$ in~$\left[0,\mathbb{I}\right]$
and $n \leq m$. Then
$\mathbb{P}_{e_k}(q) = 1$ and we get
$\mathbb{P}_{e_k}(p) = 1$ from the above assumption 
and $e_k \leq p$ from ($\ast$) for $k=1,...,n$.
Therefore, by Lemma~6~(ii), $q = \vee^{n}_{k=1} e_k \leq p$.

We now assume that $A$ is spectral 
and the state space $S_A$ is strong.
Suppose $a \in A$ with $0 \leq a \leq \mathbb{I}$ and
a minimal element $e_o \in ext(\left[0,\mathbb{I}\right])$ 
with $ \mathbb{P}_{e_o}(a) = 1 $.
Owing to the spectrality we have $a = \sum^{n}_{k=1} s_k e_k$ with $n \leq m$,
$0 \leq s_k \leq 1 $ and orthogonal
minimal extreme points $e_k \in ext(\left[0,\mathbb{I}\right])$
for $k = 1,2,...,n$.
From 
$\sum^{n}_{k=1} \mathbb{P}_{e_o}(e_k) \leq 1$ and
$1 = \mathbb{P}_{e_o}(a) = \sum^{n}_{k=1} s_k \mathbb{P}_{e_o}(e_k) $
we get 
$ \sum_{k:s_k=1} \mathbb{P}_{e_o}(e_k) = 1$.
If $\mu \in S_A$ with $\mu(e_o) = 1$, then
$\mu = \mathbb{P}_{e_o}$ by Axiom 1 and
$\mu(\sum_{k:s_k=1} e_k) = 1$.
Since $S_A$ is strong, we get 
$e_o \leq \sum_{k:s_k=1} e_k \leq a$. 
So we have verified ($\ast$).

(ii) 
Assume $q_1, q_2 \in ext(\left[0,\mathbb{I}\right])$.
We shall show that the infimum $q_1 \wedge q_2$ 
exists in $ext(\left[0,\mathbb{I}\right])$.
From $0 \leq q_1 + q_2 \leq 2 \mathbb{I} $ we get from the spectrality and Lemma 4 
$q_1 + q_2 = \sum^{n}_{k=1} s_k e_k$ with $n \leq m$,
$0 \leq s_k \leq 2$ for $k = 1,...,n$ and orthogonal
minimal extreme points $e_k \in ext(\left[0,\mathbb{I}\right])$.
For those $k$ with $s_k = 2$ then
$2 = \mathbb{P}_{e_k} (q_1 + q_2) = \mathbb{P}_{e_k} (q_1) + \mathbb{P}_{e_k}(q_2)$
and thus $\mathbb{P}_{e_k} (q_1) = 1 =\mathbb{P}_{e_k}(q_2)$;
therefore $e_k \leq q_1$ and $e_k \leq q_2$ by ($\ast$).
From Lemma~6~(ii) we get 
$\sum_{k:s_k=2} \ e_k = \vee \left\{e_k: s_k = 2\right\} \leq q_j$ for $j = 1,2$.

Now suppose $p \leq q_1$ and $p \leq q_2$ 
with $0 \neq p \in ext(\left[0,\mathbb{I}\right])$.
If $\mu(p) = 1$ for some state $\mu$, then
$\mu(q_1) = 1 = \mu(q_2)$. Therefore
$2 = \mu(q_1) + \mu(q_2) = \sum^{n}_{k=1} s_k \mu(e_k)$
and 
$\mu(\sum_{k:s_k=2} e_k) = 1$. 
Since the state space is strong, we get
$p \leq \sum_{k:s_k=2} \ e_k$.

So we have $\sum_{k:s_k=2} \ e_k = q_1 \wedge q_2$, 
and $ext(\left[0,\mathbb{I}\right])$ becomes a lattice.
For any $p, q \in ext(\left[0,\mathbb{I}\right])$
we also get the existence of the supremum $p \vee q$ via 
$(p' \wedge q')' = p \vee q$. 
The lattice $ext(\left[0,\mathbb{I}\right])$ is atomic by Lemma 3 (iv)
and orthomodular by Lemma~6~(iii).
\hfill $\square$
\newpage

The property ($\ast$) has a purely mathematical character.
Theorem 2 now shows that,
when the information capacity is finite, 
it is equivalent to
two better interpretable and physically more plausible postulates: 
spectrality and strongness of the state space.
Moreover, from Theorem 1
we can then conclude that, when Axioms 1 and 2 are given and 
the information capacity is finite, the following statements are equivalent:
\begin{itemize}
\item [-] The state space has the property ($\ast \ast$).
\item [-] The order unit space has the property ($\ast$).
\item [-] The order unit space is spectral and its state space is strong.
\end{itemize}
Any compact convex set with the property ($\ast \ast$) gives rise 
to such a structure, provided that the information capacity is finite.
The associated quantum logic becomes an atomic orthomodular lattice.
The Axioms 1 and 2 follow from ($\ast \ast$).

One customary property of the quantum logics is still missing; this
is the \emph{covering property} \cite{AS02, AIHPA1978Guz, piron1964axiomatique},
which is a technical requirement 
needed from mathematical reasons
in Piron's derivation of the Hilbert space 
from quantum logical postulates \cite{piron1964axiomatique}.
This derivation still includes non-classical Hilbert spaces
over very exotic division rings \cite{Keller1980}). 
There are a few attempts to motivate the covering property
\cite{beltrametti2010logic, cohen1987minimal, AIHPA1978Guz, moretti2019fundamental},
but no entirely convincing argument has been given
for assuming its general validity. 

When $A$ is spectral and $a \in A$, 
the square (or any real function) of $ a = \sum^{n}_{k=1} s_k e_k$
can be defined in the following way:
$$ a^{2} := \sum^{n}_{k=1} (s_k)^{2} e_k.$$
One might then be intended to construct a product on $A$ via
$$a \circ b := \frac{1}{4} \left((a + b)^{2} - (a - b)^{2}\right).$$
Generally, however, this product is not linear in $a$ or $b$.
If it is linear, we get a power-associative algebra $A$ 
and $A$ becomes a formally real Jordan algebra by a result by 
Iochum and Loupias \cite{IochumLoupias1985}.

The order unit spaces studied in the preceding paper \cite{Nie2022genqubit}
satisfy Axiom~1, are spectral, possess strong state spaces and 
the finite information capacity $m = 2$.
They give rise to the generalized qubit models.
Axiom 2 does not occur there. The reason is that Axiom 2 
becomes redundant in the case with information capacity $m = 2$. This is shown
at the beginning of the proof of part~(ii) of Theorem 9.1 in Ref. \cite{Nie2022genqubit}. 
Considering this and Theorem 9.2 in Ref. \cite{Nie2022genqubit},
we can conclude first from Theorem 3 that
the space of continuous affine functions on a 
smooth and strictly convex compact convex set has the property ($\ast$)
and then from Theorem~1 that each such set
possesses the property ($\ast \ast$),
which can also be seen directly.

\section{Symmetric transition probability}

An order unit space that satisfies the Axioms 1 and 2
possesses a \emph{symmetric} transition probability,
if $$\mathbb{P}_{e_2}(e_1) = \mathbb{P}_{e_1}(e_2)$$ 
holds for each pair $e_1,e_2$ of 
minimal extreme points in $\left[0,\mathbb{I}\right]$.
For a compact convex set $ \Omega $ with the property ($\ast \ast$)
this means $e_{\omega_1}(\omega_2) = e_{\omega_2}(\omega_1)$
for all extreme points $\omega_1$ and $\omega_2$ (see section 3).

With the order unit space formed by the self-adjoint bounded 
linear operators on a Hilbert space and one-dimensional 
orthogonal projections $e_1$ and $e_2$ we have
$$\mathbb{P}_{e_2}(e_1) 
= \left|\left\langle \eta_1|\eta_2 \right\rangle\right|^{2} 
= \left|\left\langle \eta_2|\eta_1 \right\rangle\right|^{2} 
=  \mathbb{P}_{e_1}(e_2),$$
where $\eta_1$ and $\eta_2$ are normalized elements in the 
ranges of $e_1$ and $e_2$, respectively \cite{nie2020alg_origin, nie2021generic}.
The transition probability remains symmetric
in the von Neumann algebras and JBW algebras \cite{nie2020alg_origin, nie2021generic}.

Many examples with non-symmetric transition probability
were presented in Ref. \cite{Nie2022genqubit}.
One is the \emph{triangular pillow} from Ref. \cite{AS02};
its information capacity is $m = 3$ and it
does neither possess the covering property
nor a symmetric transition probability.
The other examples have the information capacity $m = 2$;
these are the generalized qubit models.
They possess the covering property.
The associated state spaces
are the smooth and strictly convex compact convex sets. 
Here are equivalent:
\begin{itemize}
	\item [-]The transition probability is symmetric.
	\item [-]The state space is isomorphic to the unit ball in a real Hilbert space.
	\item [-]The order unit space is a so-called spin factor (this is a formally real 
or Euclidean Jordan algebra with information capacity $m=2$).
\end{itemize}
The self-adjoint real $2 \times 2$ matrices 
form the spin factor or Jordan algebra 
that belongs to the disk shown in Fig. 3.
The three-dimensional ball represents
the state space of the self-adjoint complex $2 \times 2$ matrices 
and the customary qubit.
Now any smooth and strictly convex compact convex set 
that is not isomorphic to the unit ball in some real Hilbert space 
yields an example with non-symmetric transition probability
(for instance the unit ball in any $l^{p}$ space with $1 < p < 2$ or $ 2 < p$).
\vspace{0,3 cm}

\noindent
\textbf{Theorem 3:}
\itshape
Let $A$ be an order unit space that satisfies the Axioms 1 and 2
and possesses the property ($\ast$), a finite information capacity $m$
and a symmetric transition probability.
Then $A$ becomes a real pre-Hilbert space with an inner product 
$\left\langle \  | \ \right\rangle$
such that $\left\langle e_1 | e_2 \right\rangle = \mathbb{P}_{e_1}(e_2)$
holds for all minimal extreme points $e_1$ and $e_2$ 
in $\left[0,\mathbb{I}\right]$.
Moreover, the positive cone in $A$ is self-dual; 
this means we have for any $a \in A$: $0 \leq a$ iff 
$0 \leq \left\langle a | b \right\rangle $ for all $b \in A$ with $0 \leq b$.
\normalfont
\vspace{0,3 cm}

Proof.
Define $\left\langle a | b \right\rangle := \sum^{n}_{k=1} s_k \mathbb{P}_{e_k}(b)$
for $a,b \in A$, $a = \sum^{n}_{k=1} s_k e_k$
with $s_k \in \mathbb{R}$ and 
orthogonal minimal extreme points $e_k$ in $\left[0,\mathbb{I}\right]$
($k = 1,...,n$; $n \leq m$).
Due to Theorem 2 we can assume that any $a$ and $b$ 
can be represented in this spectral form.
Obviously this inner product is linear in $b$.
Use the spectral form of $b$ and the symmetry of the transition probability,
to see that the inner product is linear in $a$ as well 
and well-defined (independent of the chosen spectral form
of $a$ which is not unique).
From $\left\langle a | a \right\rangle = \sum^{n}_{k=1} {s_k}^{2}$
we get the positive definiteness. 
Use the spectral forms of $a \geq $ and $b \geq 0$ to
show that $\left\langle a | b \right\rangle \geq 0$.
If $a \in A$ and $\left\langle a | b \right\rangle \geq 0$ 
for all $b \in A$ with $b \geq 0$,
use the spectral form $a = \sum s_k e_k$
and get $0 \leq \left\langle a | e_k \right\rangle = s_k$
for each $k$
and thus $0 \leq a$. 
Note that, by Lemma 4, $0 \leq a$ iff $0 \leq s_k$ for $k=1,2,...,n$.
\hfill $\square$
\vspace{0,3 cm}

Recall Theorem 2 and note that the property ($\ast$) in Theorem 3 
can be replaced by the condition
that $A$ is spectral and the state space $S_A$ is strong.

With the assumptions of Theorem 3 
and information capacity $m$, we have
for $a = \sum^{n}_{k=1} s_k e_k \in A$, 
$s_k \in \mathbb{R}$, 
orthogonal minimal extreme points $e_k$ in $\left[0,\mathbb{I}\right]$
and $n \leq m$
$$ \left\langle a | a \right\rangle ^{1/2} 
= \left(\sum^{n}_{k=1} s^{2}_{k}\right)^{1/2} 
\leq m^{1/2} \  max\left\{\left| s_k \right| : k=1,2,...,n \right\}
= m^{1/2} \left\|a\right\| $$
and
$$ \left\langle a | a \right\rangle ^{1/2} 
= \left(\sum^{n}_{k=1} s^{2}_{k}\right)^{1/2} 
\geq \left( max\left\{ s^{2}_{k}  : k=1,2,...,n \right\}\right)^{1/2} = \left\|a\right\| .$$
The inner product norm and the order unit norm thus become 
equivalent; this depends on the finite information capacity $m$ of $A$ 
and not on the dimension which need not be finite.

Therefore, with a finite information capacity,
$A$ becomes a Hilbert space 
with the inner product $\left\langle \ |\ \right\rangle$
if $A$ is a \emph{complete} order unit space.
The observables (including the elements of the quantum logic)
are not operators on the Hilbert space as in ordinary quantum mechanics,
but become elements in the Hilbert space.
However, we get the familiar 
duality between the states $\mu$ and the 
observables $a= \sum s_k e_k \in A$ with $0 \leq s_k $ and $\sum s_k = 1$
via $\mu(b) = \left\langle a | b \right\rangle$, $b \in A$.

The following property is often used in reconstructions of quantum mechanics, which we here transfer
to our setting: The order unit space or its state space is called \emph{strongly symmetric} if there is an
automorphism for each pair of orthogonal families of atoms (or pure states) with the same cardinality
that maps the atoms from one of the two families to the atoms (or pure states) of the other one. 
In the finite-dimensional case with symmetric transition probability, strong symmetry is possible
in our setting only if we deal with either the classical state spaces or the state spaces of the irreducible
Euclidean Jordan algebras. This follows from a result by Barnum and Hilgert \cite{BarnumHilgert2020}. Their definition of
spectrality is different, but follows from our one by the duality between the states and the observables via
the inner product. The strong symmetry is a very restrictive requirement; it rules out all the non-classical
reducible cases (direct sums).

A different approach giving rise to a self-dualizing inner product as in Theorem 3
was achieved by Wilce \cite{Wilce2019conjugatesfilters}. 

\section{A certain generalization of the Euclidean Jordan algebras}

In this section, we shall limit ourselves to the consideration
of finite-dimensional spaces. Note that a finite-dimensional 
Hilbert space over the real numbers is commonly called a \emph{Euclidean space}.

By the Koecher-Vinberg theorem \cite{Koecher1957,Vinberg1961}, 
the Euclidean spaces with \emph{homogeneous} self-dual cones 
are Jordan algebras. By this theorem,
the order unit space $A$ in Theorem~5
would become a Euclidean (or formally real) Jordan algebra 
in the finite dimensional case, if the 
positive cone were homogeneous. 

Homogeneity is a nice mathematical
property, but hard to justify as a physically indispensable postulate.
Under certain assumptions, there is relation to a quantum mechanical feature
called \emph{steering}~\cite{barnum2013ensemble}.
Furthermore, the effect algebras with the so-called sequential product 
provide a framework, where homogeneity arises quite naturally \cite{JvdW2019seq_prod_are_Jordan}.

Here we will address the question of what other properties instead of homogeneity
would impose the structure studied in sections 2 - 6
on any Euclidean space with a self-dual cone.
For this purpose we first introduce a new definition of an \emph{atom} 
in this framework.

Let $A$ with the inner product $\left\langle \  | \ \right\rangle$ 
be a Euclidean space with a self-dual cone;
this means that $A$ possesses an order relation $\leq$ with
$0 \leq a$ for any $a \in A$ iff 
$0 \leq \left\langle a | b \right\rangle $ for all $b \in A$ with $0 \leq b$.
If $a \in A$, then there are unique positive elements $a^{+}$ and $a^{-}$ in $A$
which satisfy $a = a^{+} - a^{-}$ and $ \left\langle a^{+}|a^{-}\right\rangle= 0$ \cite{penney1976self}.

Note that, in this section, \emph{orthogonality} refers to the relation 
stemming from the inner product: the elements $a,b \in A$ are orthogonal
iff $\left\langle a|b \right\rangle = 0$.

An \emph{atom} in this section is an element $0 \leq e \in A$ that satisfies
the following two conditions:
\begin{enumerate}
	\item [(i)] $e = a + b$ with $a,b \in A$, $0 \leq a,b$ and $\left\langle a|b \right\rangle = 0$ $\Rightarrow  a = 0$ or $b=0$.
	\item [(ii)] $\left\langle e|e \right\rangle = 1$.
\end{enumerate}
We shall now prove a certain spectral theorem with this type of atoms.
\vspace{0,3 cm}

\noindent
\textbf{Lemma 7:} 
\itshape
Let $A$ be a Euclidean space 
with a self-dual cone.
\begin{itemize}
	\item [(i)] If $a_1 + a_2$ and $b$ are orthogonal with $0 \leq a_1, a_2, b \in A$,
then $a_1$ and $b$ are orthogonal and $a_2$ and $b$ are orthogonal.
If $a $ and $b$ are orthogonal and $a_o \leq s a$ 
with $0 \leq a_o, a, b \in A$ and $0 \leq s \in \mathbb{R}$,
then $a_o$ and $b$ are orthogonal.
	\item [(ii)] Each $a \in A$ with $0 \leq a \neq 0$ can be represented as 
$ a = \sum^{n}_{k=1} s_k e_k$ with pairwise orthogonal atoms $e_1, ..., e_n$ and 
$0 < s_1, ..., s_n \in \mathbb{R}$ $(n \in \mathbb{N})$.
	\item [(iii)] Any $a \in A$ can be represented as 
$ a = \sum^{n}_{k=1} s_k e_k$ 
with pairwise orthogonal atoms $e_1, .... e_n$ and 
$s_1, ..., s_n \in \mathbb{R}$ $(n \in \mathbb{N})$.
\end{itemize}
\normalfont
\vspace{0,3 cm}

Proof. (i) Suppose $a_1 + a_2$ and $b$ are orthogonal with $0 \leq a_1, a_2, b \in A$.
Then $0 = \left\langle a_1 + a_2 | b\right\rangle 
= \left\langle a_1 | b\right\rangle + \left\langle a_2 | b\right\rangle$.
From $0 \leq \left\langle a_k | b\right\rangle$ for $k=1,2$
we get $\left\langle a_1 | b\right\rangle = 0 = \left\langle a_2 | b\right\rangle$. 
The second part follows with $a_1 = a_o$ and $a_2 = s a - a_o$.

(ii) Suppose $a \in A$ with $0 \leq a \neq 0$.
If $a$ cannot be written as a sum $a = b_1 + b_2$ 
with $0 \leq b_1,b_2 \in A$, $b_1 \neq 0 \neq b_2$ 
and $ \left\langle b_1|b_2 \right\rangle = 0 $,
we have $a = s_1 e_1$ with $s_1 = \left\langle a | a \right\rangle ^{1/2}$ and $e_1 = a/s_1$. 
In the other case, we consider $b_1$. If $b_1$ cannot be represented as the sum of 
two orthogonal positive non-zero elements in $A$, we have 
$a = s_1 e_1 + b_2 $ with $s_1 = \left\langle b_1 | b_1 \right\rangle ^{1/2}$ and $e_1 = b_1/s_1$.
If $b_1$ is such a sum, we consider its first summand and continue this procedure.
We thus get a sequence of non-zero positive elements in $A$
which are pairwise orthogonal by (i). 
Since the dimension of $A$ is finite, the procedure stops after a finite number of steps and  
we finally get get an atom $e_1$, $s_1 > 0$ and $0 \leq a_1 \neq 0$ 
with $a = s_1 e_1 + a_1$ and $\left\langle e_1 | a_1 \right\rangle = 0$. 

Applying the above procedure to $a_1$ instead of $a$, we get $e_2$, $s_2$ and $a_2$ with
$a = s_1 e_1 + s_2 e_2 + a_2 $ and $0 \leq a_2$,
where $e_1$, $e_2$ and $a_2$ are pairwise orthogonal. 
If $a_2 \neq 0$, 
we continue with $a_2$ and so on. Due to the finite dimension of $A$, 
we must arrive at a step where $a_{n+1} = 0$ 
and then $a = \sum^{n}_{k=1} s_k e_k$. 

(iii) Suppose $a \in A$. Then $a = a^{+} - a^{-}$ 
with $0 \leq  a^{+}, a^{-}$ and $ \left\langle a^{+}|a^{-}\right\rangle = 0$.
The case $a = 0$ is trivial, and if $a = a^{+}$ or $a = a^{+}$, 
we can immediately use part (ii).
Therefore we can assume that $a^{+} \neq 0$ and $a^{-} \neq 0$ 
and apply part (ii) to $a^{+}$ and $a^{-}$ each; 
the orthogonality of the atoms belonging to $a^{+}$ 
with those belonging to $a^{-}$ follows from (i).
\hfill $\square$
\vspace{0,3 cm}

The following property (tp) makes the system of atoms 
a so-called \emph{transition probability space}
with the inner product as transition probability. 
Transition probability spaces were 
introduced by Mielnik \cite{mielnik1968geometry} and, 
with some additional postulates, 
a "Hilbert" space over a division ring can be derived, 
but this division ring is an algebraic construct 
that need not be the real or complex numbers \cite{pulmannova1986transition}.

\begin{enumerate}
	\item [(tp)] If $e_1, e_2, ..., e_n$ is a maximal family of pairwise orthogonal atoms 
	in the Euclidean space $A$ with self-dual cone,
	then $\sum^{n}_{k=1} \left\langle e_k|e\right\rangle = 1$ for any further atom $e$.
\end{enumerate}
The following lemma provides a mathematically equivalent description of this property. 
\vspace{0,3 cm}

\noindent
\textbf{Lemma 8:} 
\itshape
Let $A$ be a Euclidean space 
with a self-dual cone. Then the following two conditions are equivalent.
\begin{enumerate}
	\item [(i)] $A$ possessess the property (tp).
	\item [(ii)] There is a positive element $\mathbb{I}$ in $A$ such that 
$\sum^{n}_{k=1} e_k = \mathbb{I}$ holds for every
maximal family of pairwise orthogonal atoms $e_1, ..., e_n$.
\end{enumerate}
\normalfont
\vspace{0,3 cm}

Proof. Assume (i) and 
let $e_1, ..., e_n$ and $f_1, ..., f_m$ each be 
a maximal family of pairwise orthogonal atoms. By (tp)
we have that $ \left\langle \sum^{n}_{k=1} e_k|e\right\rangle = 1 
= \left\langle \sum^{m}_{k=1}  f_k|e\right\rangle$ for any atom $e$.
From Lemma 7 (iii) we then get
$ \left\langle \sum^{n}_{k=1} e_k|a\right\rangle 
= \left\langle \sum^{m}_{k=1} f_k|a\right\rangle$ 
for any $a \in A$ and thus
$\sum^{n}_{k=1} e_k = \sum^{m}_{k=1} f_k$.
Now define $\mathbb{I} := \sum^{n}_{k=1} e_k$.

Assume (ii) and let $e_1, ..., e_n$ be 
a maximal family of pairwise orthogonal atoms. 
Then $\sum^{n}_{k=1} e_k = \mathbb{I}$ and 
$\sum^{n}_{k=1} \left\langle e_k | e\right\rangle 
= \left\langle \sum^{n}_{k=1}  e_k | e\right\rangle
= \left\langle \mathbb{I} | e\right\rangle
= 1$.
\hfill $\square$
\vspace{0,3 cm}

This special element $\mathbb{I}$ in $A$ will play an important role below.
Some first consequences are stated in the following lemma.
\vspace{0,3 cm}

\noindent
\textbf{Lemma 9:} 
\itshape
Let $A$ be a Euclidean space 
with a self-dual cone and the property~(tp).
\begin{itemize}
	\item [(i)] For any atom $e$ we have $ \left\langle e | \mathbb{I} \right\rangle= 1 $.
	\item [(ii)] If $a \in A$ and $0 \leq a \leq \mathbb{I}$, 
	then $ a = \sum^{n}_{k=1} s_k e_k$ with pairwise orthogonal atoms $e_1, ..., e_n$ 
and $0 \leq s_1, ..., s_n \leq 1$ $(n \in \mathbb{N})$.
	\item [(iii)] The special element $\mathbb{I}$ is an order unit.
	Note that the norm arising from the order unit is not the same as the one 
	arising from the inner product.
\end{itemize}
\normalfont
\vspace{0,3 cm}

Proof. (i) Choose a maximal family of pairwise orthogonal atoms including $e$; 
since its sum is $\mathbb{I}$, we get $ \left\langle e | \mathbb{I} \right\rangle= 1 $.

(ii) Suppose $0 \leq a \leq \mathbb{I}$. By Lemma 7 (ii), 
$ a = \sum^{n}_{k=1} s_k e_k$ with pairwise orthogonal atoms $e_k$ 
and $0 \leq s_k $. Then $s_k = \left\langle e_k|a\right\rangle 
\leq \left\langle e_k| \mathbb{I} \right\rangle = 1$ for each $k$.

(iii) Let $a$ be any element in $A$. By Lemma 7 (iii), it can
be represented as 
$ a = \sum^{n}_{k=1} s_k e_k$ with pairwise orthogonal atoms $e_k$ and
$s_k \in \mathbb{R}$ $(n \in \mathbb{N}, k = 1,2,...,n)$.
With $s := max\left\{\left|s_1\right|, \left|s_2\right|, ..., \left|s_n\right|\right\}$ 
we then have 
$$-s \mathbb{I} \leq -s\sum^{n}_{k=1} e_k \leq a \leq s\sum^{n}_{k=1} e_k \leq s \mathbb{I}.$$ 
Moreover, 
if $m a \leq \mathbb{I}$ holds for all $m \in \mathbb{N}$, 
we get for any $b \in A $ with $0 \leq b$: 
$m \left\langle b | a \right\rangle \leq \left\langle b | \mathbb{I} \right\rangle $ 
for all $m$
and thus $\left\langle b | a \right\rangle \leq 0$. Therefore $b \leq 0$.
This means that $\mathbb{I}$ is an order unit 
and $A$ becomes an order unit space \cite{AS01, AS02, hanche1984jordan}.
\hfill $\square$
\vspace{0,3 cm}

The following property represents an analog of ($\ast$) and
will finally ensure that $A$ possesses
the mathematical structure presented in section 2.

\begin{enumerate}
	\item [($\ast \ast \ast$)] If $\left\langle e|a \right\rangle = 1$ for some atom $e$ in $A$
and some $a \in A$ with $0 \leq a \leq \mathbb{I}$, then $e \leq a$.
\end{enumerate}

\noindent
\textbf{Theorem 4:} 
\itshape
Let $A$ be a Euclidean space 
with a self-dual cone and the properties (tp) and ($\ast \ast \ast$).
\begin{itemize}
	\item [(i)]  $A$ 
	satisfies the Axioms 1 and 2 as well as ($\ast$). 
	The transition probability is symmetric and coincides with the inner product.
	\item[(ii)] The state space is
	$\left\{a \in A | 0 \leq a \text{ and } \left\langle a | \mathbb{I} \right\rangle = 1\right\}$
	and possesses the property ($\ast \ast$).
\end{itemize}
\normalfont

Proof. (i) First we investigate the extreme points of $\left[0,\mathbb{I}\right]$.
Suppose $p \in ext(\left[0,\mathbb{I}\right])$. 
From Lemma 9 (ii) we get $p = s_1 e_1 + ... s_n e_n$ 
with pairwise orthogonal atoms $e_1, ..., e_n$ and $0 \leq s_k \leq 1$ for each $k$.
If $0 < s_j < 1$ for some $j$, then 
$p = (1 - s_j) (\sum_{k \neq j} s_k e_k) + s_j (e_j + \sum_{k \neq j} s_k e_k$,
which means that $p$ is not an extreme point.
Therefore the extreme points of $\left[0,\mathbb{I}\right]$ 
have the form $p = e_1 + ... e_n$ with pairwise orthogonal atoms $e_1, ..., e_n$. 

We now show that the atoms as defined in this section coincide with the
minimal extreme points in $\left[0,\mathbb{I}\right]$. 
Let $e$ be such an atom.
If $e = ta + (1-t)b$ with $0 < t < 1$ and $a,b \in \left[0,\mathbb{I}\right]$, then 
$1 = \left\langle e|e\right\rangle 
= t \left\langle e|a \right\rangle+ (1-t) \left\langle e|b\right\rangle $ and
from $\left\langle e|a \right\rangle \leq \left\langle e|\mathbb{I}\right\rangle = 1$
and $\left\langle e|b \right\rangle \leq \left\langle e|\mathbb{I}\right\rangle = 1$
we get $\left\langle e|a \right\rangle = 1 = \left\langle e|b \right\rangle$.
The property ($\ast \ast \ast $) implies $e \leq a$ and $e \leq b$.
For all positive $x \in A$ we then have
$\left\langle e|x\right\rangle \leq \left\langle a|x\right\rangle$, 
$\left\langle e|x\right\rangle \leq \left\langle b|x\right\rangle $
and
$\left\langle e|x\right\rangle 
= t \left\langle a|x\right\rangle + (1-t) \left\langle b|x\right\rangle$;
therefore $\left\langle e|x\right\rangle 
= \left\langle a|x\right\rangle = \left\langle b|x\right\rangle$.
Since the positive elements generate $A$, we finally get
$a = b = e$ and we have shown that $e \in ext \left[0,\mathbb{I}\right]$.

Now suppose that there is $p \in ext \left[0,\mathbb{I}\right]$ with $p \leq e$.
At the beginning of this proof we have shown that
$p = e_1 + ... e_n$ with pairwise orthogonal atoms $e_1, ..., e_n$.
Then $1 = \left\langle e_k|p \right\rangle \leq \left\langle e_k|e \right\rangle \leq 1$
and thus $\left\langle e_k|e \right\rangle = 1$ for $k=1, ..., n$.
Because of $\left\langle e|e \right\rangle = 1 = \left\langle e_k|e_k \right\rangle$
this implies $e_k = e$ for each $k$,
and the orthogonality of $e_1, ..., e_n$ requires $n=1$.
Therefore $p = e$ and we have shown that $e$ is a minimal extreme point.

It remains to verify that each minimal extreme point $p$ is an atom.
We know that $p = e_1 + ... e_n$ with pairwise orthogonal atoms $e_1, ..., e_n$
and that these atoms are non-zero extreme points. The minimality of $p$ 
requires that $n=1$ and $p = e_1$ is an atom.

Because of the self-duality the state space of the order unit space $A$
is $S_A = \left\{a \in A | 0 \leq a \text{ and } \left\langle a | \mathbb{I} \right\rangle = 1 \right\}$;
the state $A \ni x \rightarrow \left\langle a|x \right\rangle $ belongs to $a \in S_A$.
We shall now check Axiom 1, Axiom 2 and the property ($\ast$)

Axiom 1: Let $e$ be a minimal extreme point. That means $e$ is an atom and we have
$\left\langle e|e\right\rangle = 1$. 
Now assume $\left\langle a|e\right\rangle = 1$ 
for some $a \in S_A$. By Lemma 7 (ii) 
$a = s_1 e_1 + ... + s_n e_n$ with pairwise orthogonal atoms $e_k$ and $0 < s_k$.
Then
$1 = \left\langle a | \mathbb{I} \right\rangle = s_1 + ... + s_n$
and 
$1 = \left\langle a|e\right\rangle 
= s_1 \left\langle e_1|e\right\rangle + ... + s_n \left\langle e_n|e\right\rangle$.
Since $0 \leq \left\langle e_k|e\right\rangle \leq 1$ for each $k$, this requires that
$\left\langle e_k|e\right\rangle = 1$ for each $k$. 
Because of $\left\langle e|e \right\rangle = 1 = \left\langle e_k|e_k \right\rangle$
this implies $e_k = e$ for each $k$.
From the orthogonality of the $e_k$ we get $n=1$ and $a=e$.

Axiom 2: We have to show the the extreme points of $S_A$ are the atoms.
Suppose $a$ is an extreme point in $S_A$. In the same way as in the above check of Axiom 1 we get again
$a = s_1 e_1 + ... + s_n e_n$ with pairwise orthogonal atoms $e_k$, $0 < s_k$
and $1 = s_1 + ... + s_n$. Thus $a$ becomes a non-trivial convex combination of the 
$e_k \in S_A$ unless $n = 1$ and $a = e_1$.
Therefore $a$ is an atom.

Now consider an atom $e$ and suppose $e = t a + (1-t) b$ 
with $a,b \in S_A$ and $0 < t < 1 $. Then
$1 = \left\langle e|e\right\rangle 
= t \left\langle e|a\right\rangle + (1-t) \left\langle e|b\right\rangle$
and $\left\langle e|a\right\rangle \leq \left\langle \mathbb{I}|a\right\rangle = 1$, 
$\left\langle e|b\right\rangle \leq \left\langle \mathbb{I}|b\right\rangle = 1$.
Therefore $\left\langle e|a\right\rangle = 1 = \left\langle e|b\right\rangle$.
Again $a = s_1 e_1 + ... + s_n e_n$ with pairwise orthogonal atoms $e_k$, $0 < s_k$
and $1 = s_1 + ... + s_n$. Thus 
$1 = \left\langle e|a\right\rangle
= s_1 \left\langle e|e_1\right\rangle + ... + s_n \left\langle e|e_n\right\rangle$
with $0 \leq \left\langle e|e_k \right\rangle \leq 1$ for each $k$.
This requires $\left\langle e|e_k \right\rangle = 1$ for each $k$
and, because of $\left\langle e|e \right\rangle = 1 = \left\langle e_k|e_k \right\rangle$,
this implies $e_k = e$ for each $k$.
From the orthogonality of the $e_k$ we again get $n=1$ and $a=e$. 
In the same way it follows that $b = e$ and we have verified that the atom $e$
is an extreme point in $S_A$.

The property ($\ast$) immediately follows from ($\ast \ast \ast$), 
and part (ii) follows from Theorem 1 (i).
\hfill $\square$
\vspace{0,3 cm}

Note that 
$\left\langle a|\mathbb{I} \right\rangle= \left\langle \mathbb{I}|a\right\rangle $
represents something very similar to the trace of the element $a \in A$ 
and coincides with the trace if $A$ is a Euclidean Jordan algebra.

By Theorem 4 the two properties (tp) and ($\ast \ast \ast $)
of a Euclidean space with self-dual cone
are sufficient to give it the mathematical structure 
presented in the sections 2 - 6 and we get many 
quantum theoretical features:
\begin{itemize}
	\item [-] a quantum logic that is an atomic orthomodular lattice,
	\item [-] a spectral decomposition for the observables,
	\item [-] a symmetric transition probability for each pair of atoms and 
	\item [-] the duality between states and positive observables with normalized "trace".
\end{itemize}
Here the observables are elements in the (real) Hilbert space,
while they are operators on it in customary quantum mechanics.

The Euclidean (formally real) Jordan algebras provide many examples 
for the Euclidean spaces with self-dual cones 
that satisfy (tp) and ($\ast \ast \ast $). Further examples are unknown
and the questions arise whether further ones exist, 
whether they can be classified in a similar way 
as the Euclidean Jordan algebras~\cite{hanche1984jordan, von1933algebraic}
and what their symmetry groups are.

\section{Discussion}

We have seen in sections 6 and 7 that 
we come close to the formally real (Euclidean) Jordan algebras
in the finite dimensional case with symmetric transition probability.
We cannot reconstruct quantum mechanics or the Jordan algebras, 
but this might be an opportunity
to detect physically meaningful more general theories.
The mathematical structure presented here 
establishes an extension of the Euclidean Jordan algebras.
Considering the close relation between them
and Lie groups \cite{baez2002octonions}, 
there may be a chance that this extended structure
is related, in some cases, to those exceptional
Lie groups where no Euclidean Jordan algebra
is associated ($E_6, E_7, E_8$).
These groups are used in string theory \cite{a1987superstring}.
Could there be any compact convex sets with the property ($\ast \ast$)
and an $E_6, E_7$, or $E_8$ symmetry? 
Or could any Euclidean space that possesses a self-dual cone
and satisfies (tp) and ($\ast \ast \ast $)
have such a symmetry?

What would it then mean to move away 
from the ordinary Hilbert space (or operator algebra) formalism of quantum mechanics to
the mathematical structure presented here, 
but to retain the symmetry of the transition probability? 
Familiar features that would remain are
the spectral decomposition of the observables 
(needed to define their probability distributions in a given state)
and the duality between the states and the 
observables $\sum s_k e_k$ with 
$0 \leq s_k \in \mathbb{R}$, $\sum s_k = 1$
and $0 \leq e_k$, $\sum e_k \leq \mathbb{I}$ (see section 6).
Moreover, the quantum logic would still be 
an atomic orthomodular lattice 
albeit without the covering property.

The generator of a continuous  (Hamilton operator) 
would be a derivation (operator on $A$), but not an observable
(element in $A$). So the physical quantity \emph{energy} 
could not any more be represented by an observable 
as in common quantum mechanics. This is the same situation as in a
Jordan algebra that is not the self-adjoint part of 
some complex $C^{\ast}$-algebra (see \cite{AS02} Theorem 6.15).
The condition that the dynamic evolution is generated 
by an observable is very closely related to the need
of the complex numbers in quantum theory \cite{barnum2014higher, nie2015dyn}.

There would be no locally tomographic mathematical model for a multipartite system,
since such a tensor product is not generally available 
just as it does not generally exist for the 
formally real Jordan algebras. Only if the Jordan algebras
are the self-adjoint parts of complex $C^{\ast}$-algebras,
they can be combined in a locally tomographic tensor product~\cite{nie2020loc_tomography}. 
Nonetheless logically independent and compatible sublogics
of the associated quantum logic can be considered 
in the general case; these sublogics almost behave 
like the components of a tensor product~\cite{nie2020loc_tomography, nie2021generic}.

A general product for the observables would not exist.
We might not even have the compressions, filters or extended conditional probabilities,
which are usually needed to get an 
equivalent of the L\"uders - von Neumann
measurement process \cite{AS02, barnum2014higher, foulis1992filters, 
AIHPA1978Guz, mielnik1969theory, niestegge2008approach, nie2012AMP},
and such an equivalent would not be available any more.
When the measurement outcome is represented
by an atom $e$ in the quantum logic, the state after the measurement
becomes $\mathbb{P}_e$ regardless of the initial state before the measurement.
However, when the measurement outcome is represented
by a non-atomic element $p$ in the quantum logic, 
it would not generally be possible to derive a post-measurement state 
from the pre-measurement state.
A paradigm shift would be required, since it could not anymore be
taken for granted that each physical system is in a quantum state.

The calculation of post-measurement probabilities 
from a pre-measurement state
would remain possible 
for those observables $a$ that lie together with the non-atomic $p$ in 
a subspace that forms a Jordan algebra with the product
$\circ$ considered in section~5. The map 
$a \rightarrow \left\{p,a,p\right\} := 2 p \circ \left(p \circ a\right) - p \circ a$
would then become a compression on this subspace and could be used to 
calculate the post-measurement probabilities from the pre-measurement state 
for those observables $a$ that lie in this subspace
\cite{AS02, niestegge2008approach, nie2012AMP, nie2020charJordan}.
In other cases, this might result in a new level of incompatibility
for $p$ and those observables for which 
the calculation of post-measurement probabilities 
from the measurement outcome $p$ and a pre-measurement state 
is not generally possible. This would go far beyond the
usual quantum mechanical incompatibility (non-commuting observables). 

Furthermore, some mathematical or physical constructs, which are 
sometimes introduced as principles in quantum mechanical reconstruction efforts,
require the compressions or their analogues and
cannot be transferred to the setting of this paper.
These are for instance:
third order interference and its absence 
\cite{barnum2014higher, nie2012AMP, nie2020charJordan}, 
ellipticity \cite{AS02}
and a certain symmetry condition (9.29 and 9.40 in \cite{AS02}, (A1) in \cite{niestegge2008approach}).

\section{Conclusions}

Here we have presented three different accesses 
to a mathematical structure with the
desired transition probability; their
starting points are (1) order unit spaces,
(2) convex sets and (3) self-dual cones.

(1) The first access begins with
a quantum logic, the associated state space
and the significant postulate that
the transitions probabilities of the quantum logical atoms
do exist (Axiom 1). It is assumed that the quantum logic
is represented by the extreme boundary 
of the unit interval in an order unit space.
Further postulates are Axiom 2 and the property ($ \ast $).
These assumptions include the customary Hilbert space quantum logic
of usual quantum theory and the projection lattices
in the operator algebras.

(2) A single geometric property 
that makes any finite dimensional compact convex set 
a matching state space was identified.
This property ($ \ast \ast $) may be hard to verify for a given set,
but this is not worse than with Alfsen and Shultz's theory \cite{AS02},
which involves some complicated mathematical constructions 
like the projective faces, compressions and projective units. 
The compressions are related to
the filters used in other approaches \cite{AIHPA1978Guz, mielnik1969theory}
and to the extended conditional probabilities introduced 
by the author \cite{niestegge2008approach, nie2012AMP, nie2020charJordan},
but all these constructions 
are not used here and need not exist generally
(However, they do exist in all the examples considered here and 
it would be interesting to identify an example where they do not exist).
Our elementary approach is more general,
particularly in the cases with finite information capacity.
The geometric property then
becomes equivalent to two desirable features 
(spectrality and strongness of the state space)
and implies that the quantum logic
is an atomic orthomodular lattice.

(3) With symmetric transition probability,
the approach presented here
results in Euclidean spaces with self-dual cones,
which can also be used as the starting point.
With the properties (tp) and ($\ast \ast \ast$),
they provide 
a certain generalization of the 
Euclidean Jordan algebras and
can be considered a mathematical model for an
extension of quantum theory.
Some familiar features remain, 
others get lost as they do already in the Jordan algebra setting.
Interesting new implications on physics concern 
the theoretical foundations of
quantum measurement and a new level of incompatibility,
which were discussed in section 8.

There is a natural candidate for a product which,
however, becomes a reasonable bilinear product
on the entire space iff this space is a formally real Jordan algebra.
A further missing property
is the covering property from which 
the dimension function can be derived.
The symmetry of the transition probabilities also
yields the dimension function \cite{nie2021generic, Nie2022genqubit, pulmannova1986transition}, 
from which, however, the stronger covering property cannot be recovered.

\bibliographystyle{abbrv}
\bibliography{Literatur2024}
\end{document}